# Optimisation and Precision Tuning of Localised Surface Plasmon Resonance in AuFON Systems

Luis Alfonso Guerra Hernández[1*], Osmar Gil Salas[1], Jorge Enrique Rueda Parada[1], Alejandro Fainstein[2,3] and Andrés Alejandro Reynoso[2,3]

[1] Grupo de Investigación Óptica Moderna (GOM)-Minciencias, Departamento de Física, Facultad de Ciencias Básicas, Universidad de Pamplona-Colombia
[2] Centro Atómico Bariloche e Instituto Balseiro, Comisión Nacional de Energía Atómica (CNEA) - Universidad Nacional de Cuyo (UNCUYO), 8400 Bariloche, Argentina
[3] Instituto de Nanociencia y Nanotecnología (INN-Bariloche), Consejo Nacional de Investigaciones Científicas y Técnicas (CONICET), Argentina

**Abstract.** Metal film on nanosphere (MFON) plasmonic systems have emerged as nanostructures with useful properties for molecular detection. This work presents the optimisation and discussion of the characterisation results of Au-film on nanosphere (AuFON) systems. Through experiments and numerical simulations, the resonance energies of localised surface plasmon modes and the spatial distribution of light on the nanostructured surface were identified. The results highlight the dependence of these modes on changes in nanostructure dimensions and the optical conditions of the incident radiation, thereby optimising the amplification of optical signals for applications in molecular detection.

## 1. Introduction

Plasmonic properties in metallic nanostructures have been of great interest due to their ability to selectively absorb and scatter photons [1-7]. In particular, these nanostructures exhibit localised surface plasmon resonances (LSPR), with electromagnetic fields confined to the surface and a resonance energy dependent on the metal, size, and shape of the nanostructure [8-10]. This interaction between photons and localised plasmons leads to the confinement of the electromagnetic field in sub-wavelength regions, enabling processes such as surface-enhanced Raman spectroscopy (SERS), among others [11-16]. For these purposes, metallic nanoparticle systems have been the most widely used platforms, as SERS enhancements are associated with electromagnetic hot spots between the nanoparticles, with enhancement factors in the range of $10^{11}$ in small surface areas [17]. However, reproducibility over large surfaces has been a challenge, which has been addressed using more ordered platforms, such as nanosphere lithography (NSL) arrays [18-22]. This technique involves assembling nanometrically precise ordered arrays of polymer nanospheres on smooth surfaces [18,19,21,23,24]. These assemblies can be used as lithographic masks for metal deposition, creating devices for SERS applications [13,14,25-27]. Among these devices, triangular nanoparticles [18,28-31], metallic sphere-segment void (SSV) cavities [25,32-35], and metal film on nanosphere (MFON) substrates [36-39] stand out. MFON systems have been proposed for various applications, such as biosensors [40-47], single-molecule detection [48-50], explosive detection [51], and art conservation [24,52,53], among others.

LSPR in MFON systems can be tuned by varying the nanosphere diameter, with applications in molecular detection across a wavelength range of ~400-1100 nm [24], and SERS amplification factors reported in the literature of up to ~$10^8$ [30,37,54-56]. These factors have improved with the formation of nanopillars through rotation during metal deposition [39]. Recently, Dmitry K. and collaborators [57] discussed the relationship between far-field and near-field properties with SERS amplifications in these systems, showing that the spectral position of the resonance depends on the surface morphology and roughness. Other studies optimised these systems for excitation at 1024 nm, achieving SERS improvements in the near-IR region [58]. The modification of Ag surfaces with alumina layers allowed their stabilisation and functionalisation for SERS applications [59], preserving activity and controlling oxidation, which doubles sensitivity and reduces measurement times. These precedents demonstrate that MFON systems have been widely used for molecular detection applications. However, these systems are often used without proper optical optimisation, with the incident radiation out of resonance with the localised plasmons of the nanostructure, limiting the efficiency of optical amplification. Therefore, a comprehensive characterisation is needed to identify the energy position where the plasmons of the metallic nanostructure resonate. This work addresses this optimisation, presenting a detailed characterisation of a Au-film on nanosphere (AuFON) plasmonic system, which improves the optical configuration for technological applications involving plasmon resonance. The results include the morphological characterisation of the sample, experimental conditions, and optical reflectivity simulations. Finally, the plasmonic responses are discussed as a function of wavelength, incident radiation polarisation, and nanostructure diameter.

* Corresponding author: luisguerra@unipamplona.edu.co

## 2. Experimental methods and simulations

### 2.1 Fabrication of plasmonic nanostructures

For the fabrication of the AuFON substrates, polystyrene nanospheres (500–800 nm in diameter, with a standard deviation ranging from 6 to 9 nm depending on the nominal size) were dispersed in a 1% aqueous solution (Duke Scientific) and assembled on a glass plate coated with a 100 nm Au film (*Platypus*). Following this self-assembly, an additional Au layer was deposited onto the polystyrene nanospheres via physical vapour deposition (Vega y Camji S.A.I.C.) in a homemade evaporator, which reached a residual pressure of ~$1x10^{-7}$ Torr. The melting temperature of the Au source was manually controlled using a current of 10–15 A. For the samples used in the experiments, the metallic layer deposited on the nanosphere array had a thickness of approximately 140 nm. A more detailed description of the AuFON substrate fabrication process can be found in Ref. 60.

### 2.2 Optical Reflectivity Measurements

Figure 1a shows a schematic of the optical reflectivity experiment, illustrating the interaction of light with the AuFON nanostructured system studied in this work. Reflectivity measurements were used to identify the plasmonic modes of the AuFON arrays. A fully automated variable-angle spectroscopic ellipsometer (Woollam WVASE32) with focusing probes was used for this purpose. The setup featured a 100 μm circular spot on the sample and a numerical aperture of ~0.02.

### 2.3 Scanning Electron Microscopy (SEM)

Figure 1b shows a high-resolution SEM image of a typical AuFON nanostructure, in which a good arrangement of the nanospheres can be observed. For the samples used in the experiments, the metallic layer deposited on the nanosphere array had a thickness of approximately 140 nm. The images were obtained using a FEI field emission SEM, model Nova NanoSEM 230, operating at 10 kV and with a tilt angle of 45°.

### 2.4 Simulation of Plasmonic Resonances in AuFON Nanostructures

Figure 1c shows the unit cell of the geometry used in the simulation of the AuFON plasmonic system. This cell consists of half a gold nanosphere supported on a flat gold surface with a thickness of 100 nm (h). The electromagnetic response of the system was obtained using a plane wave, with TM or TE polarisation. The incidence (θ) and azimuthal (ϕ) angles were varied from 0° to 90° in 5° steps. The variation of ϕ is related to the rotation of the sample in the x-y plane, around the z-axis. The unit cell dimensions were also modified by varying the nanosphere diameter between 500 and 900 nm. The dielectric function of the metal was obtained from the data of Johnson and Christy [61]. The simulation was performed using the finite element method with COMSOL Multiphysics software [62], solving Maxwell's equations in 3D for each mesh or fraction of the nanostructure.

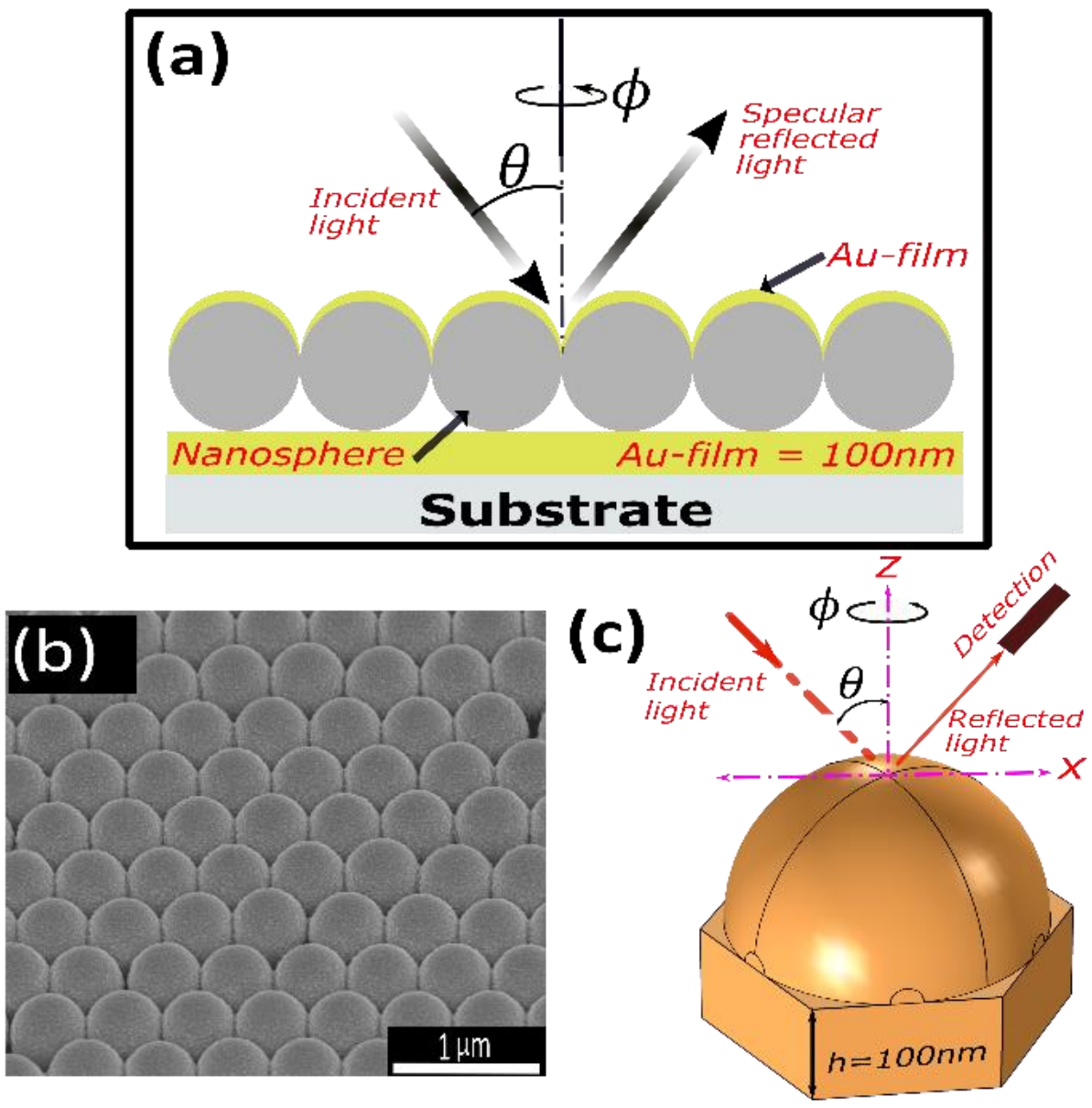

**Fig. 1.** (a) Schematic of the optical reflectivity, where the incident light interacts with an AuFON nanostructured system. (b) High-resolution SEM image of a typical AuFON nanostructure with polystyrene nanospheres of 500 nm in diameter. (c) Unit cell of the simulated nanostructured system. The plane wave with TM or TE polarisation is incident on the metallic nanostructure. The incidence (θ) and azimuthal (ϕ) angles can be varied in the simulation, as well as the dimensions of the nanostructure.

## 3. Results and Discussion

The characterisation of the spectral response of the plasmonic modes of AuFON systems was carried out through optical reflectivity measurements and the evaluation of the average electric field amplitude on the nanostructure surface. The discussion of the results focuses on establishing the resonance energy of the plasmonic modes and their dependence on the incident light polarisation, the nanostructure diameter, and the angles θ and ϕ. We begin discussing the wavelength dependence of the reflectivity. Figure 2 presents the spectral response results for two AuFON systems, with diameters of 500 and 600 nm, for TM and TE polarised light. In both cases (experiment and simulation), an incidence angle of 25° and an azimuthal angle of 0° were used. The results show a strong agreement between the experiment and the simulation. In each panel, two intense absorptions are identified in the wavelength range of 600-900 nm, corresponding to localised surface plasmon resonances. For the 600 nm diameter sample, the resonances are located at 664 nm (LSPR1) and 830 nm (LSPR2) for TM polarisation, and at 660 nm (LSPR1) and 808 nm (LSPR2) for TE polarisation. For the 500 nm sample, the plasmonic resonances are found at 591 nm (LSPR1) and 740 nm (LSPR2) for TM polarisation, and at 574 nm (LSPR1) and 700 nm (LSPR2) for TE polarisation. At wavelengths < 500 nm, a decrease in reflectivity is observed due to the interband transitions characteristic of Au [63]. Figure 3 shows the results of the simulation of the plasmonic response for a 500 nm diameter AuFON system, for TM polarisation (panel (a)) and TE polarisation (panel (b)). Each reflectivity curve was obtained with an incidence angle θ=25°, and the azimuthal angle ϕ was varied from 0° to 90° in 10° increments. The results show that, for both polarisations (TM and TE), the resonance energy of the plasmonic modes (LSPR1 and LSPR2) does not significantly depend on ϕ, which can be attributed to the honeycomb ordering of the nanostructures with high symmetry respect to rotation in the xy-plane. This same behaviour was observed for all nanostructures modelled irrespective of the nanosphere diameter (500-900 nm).

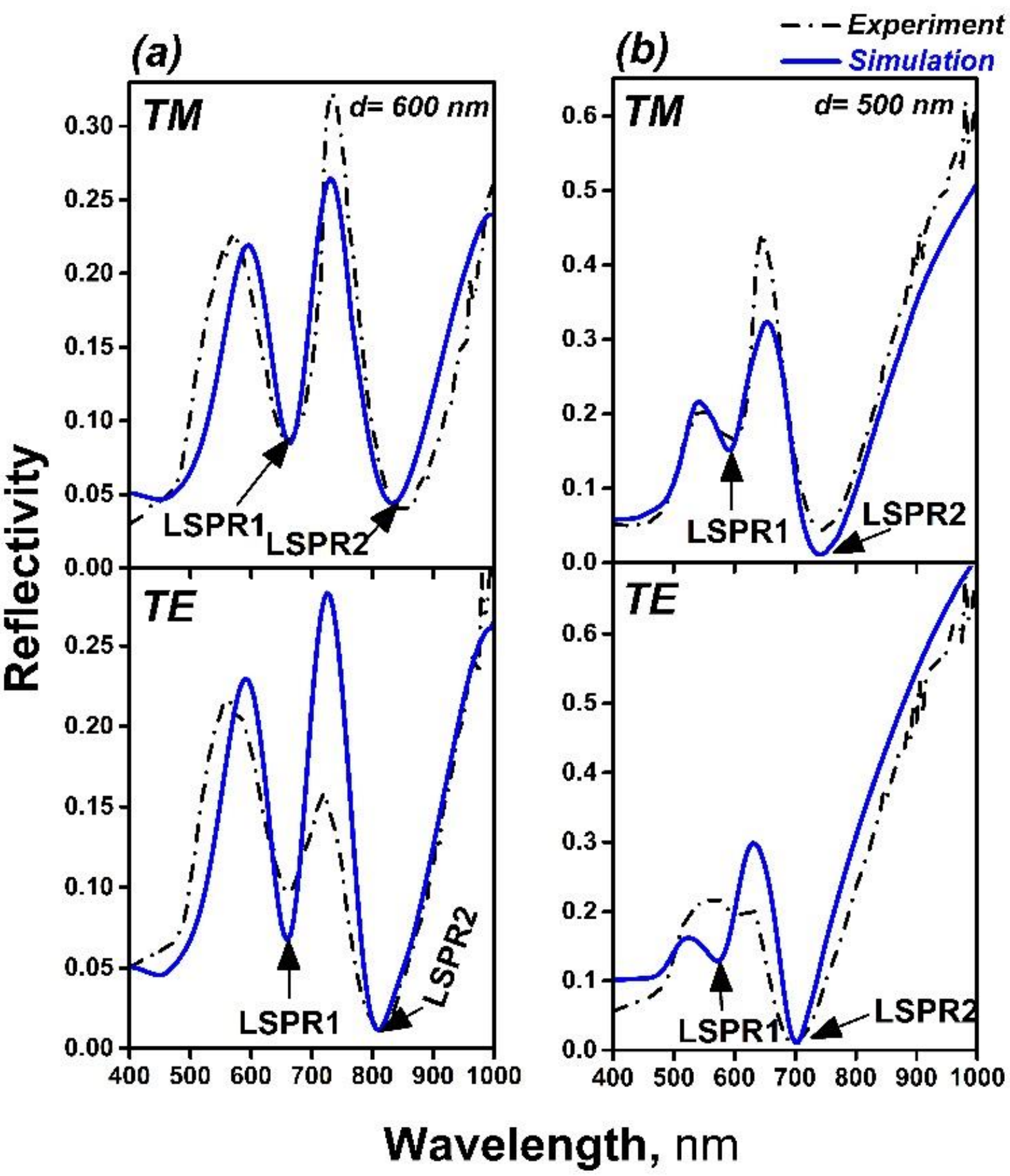

**Fig. 2.** Simulation and experimental results of the spectral response of localised plasmonic modes in AuFON systems for TM (top) and TE (bottom) polarisations; in (a) for a 600 nm diameter sample and in (b) for a 500 nm diameter sample. The incidence angle (θ) was 25°, and the azimuthal angle (ϕ) was 0°.

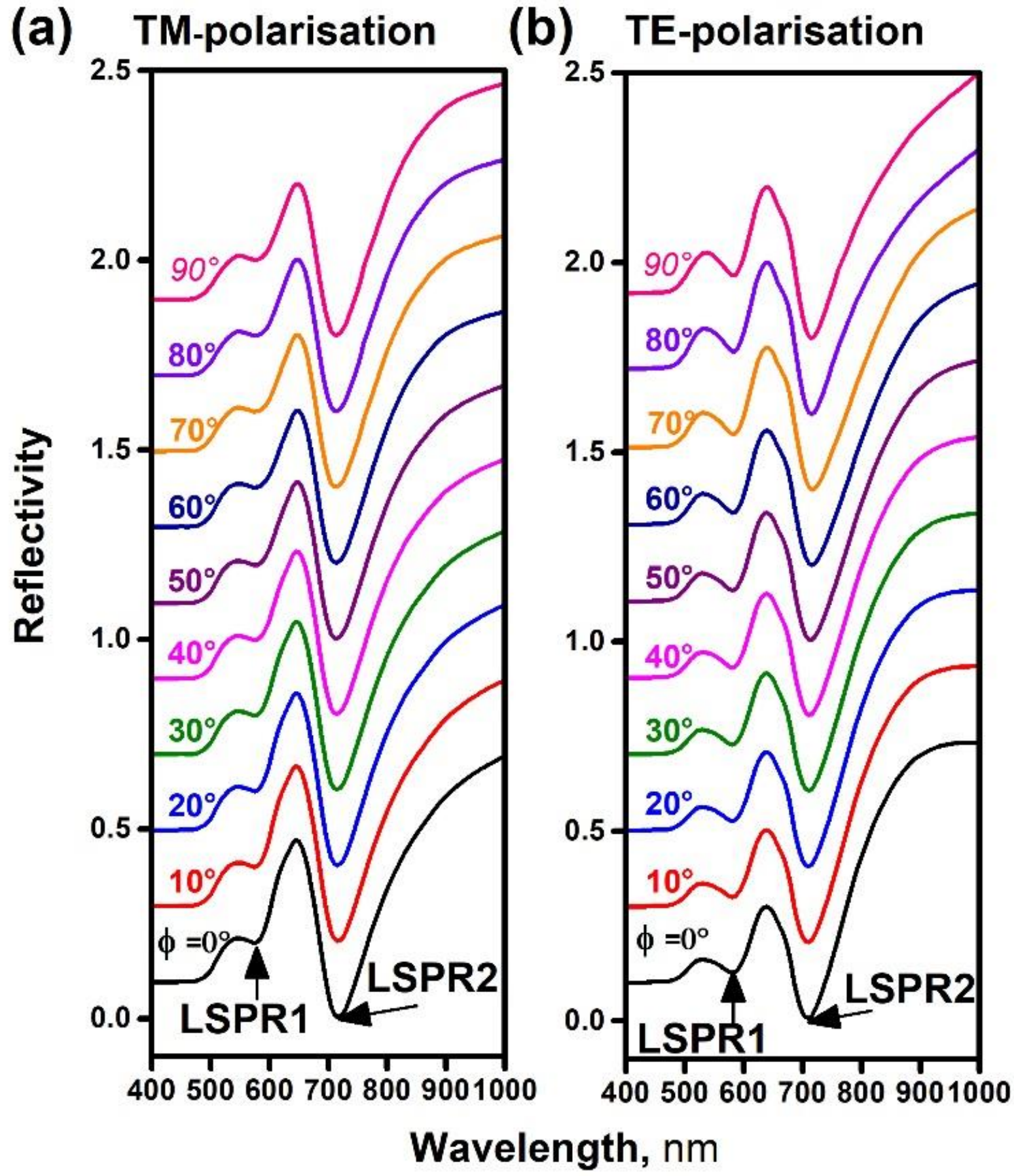

**Fig. 3.** Simulation of the plasmonic response for a 500 nm diameter AuFON sample, for TM polarisation (panel (a)) and TE polarisation (panel (b)). The incidence angle was fixed at θ=25°, and ϕ was varied from 0° to 90° in 10° increments. The curves are vertically shifted by 0.2 for clarity.

The in-plane dependence with the angle ϕ of the plasmonic response for a 500 nm diameter sample under TM polarisation (panel (a)) and TE polarisation (panel (b)) are presented in Figure 4. The angle of incidence was set to 25°, while the angle ϕ was varied from 0° to 90°. In agreement with the numerical simulations presented in Figure 3, the experimental resonance energy of the localised plasmonic modes (LSPR1 and LSPR2) does not exhibit significant changes on varying ϕ. These results confirm the good ordering of the synthesised plasmonic nanostructures, and suggest their ease of use in applications.

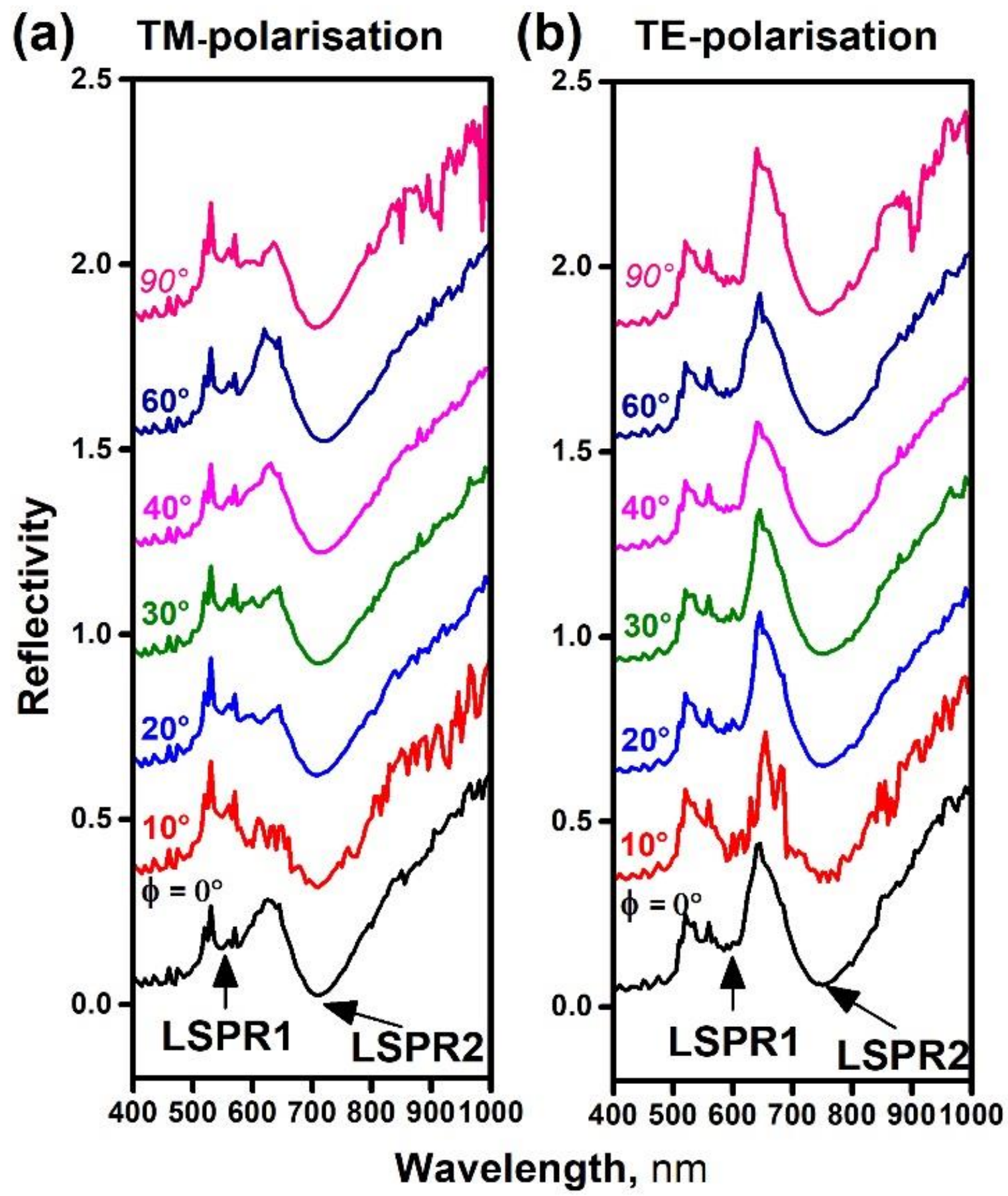

**Fig. 4.** Experimental results of the plasmonic response for a 500 nm diameter AuFON sample, for TM polarisation (panel (a)) and TE polarisation (panel (b)). The angle of incidence was fixed at 25°, and ϕ was varied from 0° to 90°. Each reflectivity curve was shifted vertically for clarity.

Figure 5 shows simulated results of the spectral response of a 500 nm diameter AuFON nanostructured system, with TE polarisation and different values of θ. We note that the modelling for TM polarisation shows a very similar response, and lack of variation with θ. In panel (a), the result for optical reflectivity is presented, where two resonant modes (LSPR1 and LSPR2) are identified. These modes show a low dependence of their resonance energy on the incidence angle θ. In panel (b), the average amplitude of the electric field associated with each reflectivity curve is shown. The spectral position of the plasmons (LSPR1 and LSPR2) in reflectivity coincides with the maximum of the average electric field amplitude, with LSPR2 always having the highest intensification. This indicates that the LSPR2 mode has a higher capacity for light concentration in the metallic nanostructure, favouring its use in amplification applications for detection. Figure 6 shows experimental results of the plasmonic response in a 500 nm diameter AuFON sample, analysed under different angles of incidence (θ) with a constant azimuthal angle (ϕ). Panels (a) and (b) show the reflectivity corresponding to the TE and TM polarisations, respectively. For TE polarisation (panel (a)) the agreement with the computational simulations (Figure 5a) is excellent, both displaying the same spectral shape and independence with variations of θ. For the TM polarisation spectra (panel (b)), the experimental behaviour is more complex. The mode labelled LSPR1 seems to remain independent with θ, in agreement with calculations. However, an angular dispersion with a gradual red shift with increasing θ is observed for the mode identified as LSPR2, and also for an additional feature that seems to split from LSPR1. Such dispersive behaviour is typical of propagating Bragg modes as also observed in periodic nanovoids of shallow thickness [64]. The coexistence of localised plasmonic modes and Bragg-type modes in periodic arrays of truncated spherical cavities was studied. It was observed that localised modes dominate under certain geometric conditions of the nanostructure, but as the degree of periodicity increases, propagating Bragg modes begin to couple with localised modes, resulting in a more complex angular dispersion. Although the AuFON systems differ geometrically from those studied in that reference, a conceptual analogy can be established. In particular, the periodic arrangement of the nanospheres in our substrates may favour coupling between localised and propagating modes, which could explain the red shift observed in LSPR2 with increasing angle θ.

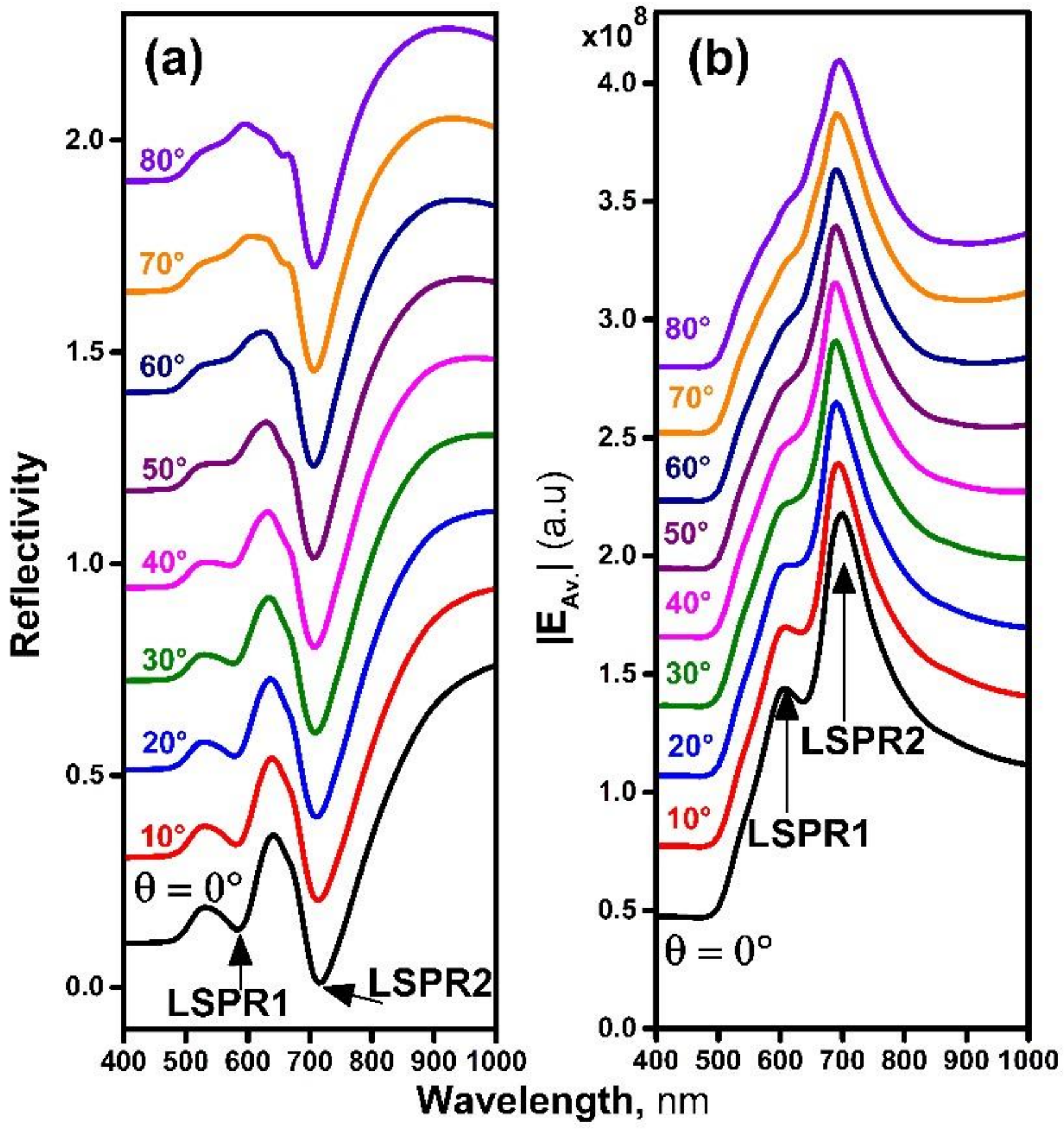

**Fig. 5.** Simulated spectral response for a 500 nm diameter AuFON sample with different values of θ. (a) Optical reflectivity, and (b) the associated average electric field amplitude. For these results, ϕ=25° and TE polarisation were used. The curves are vertically shifted by 0.3 for clarity.

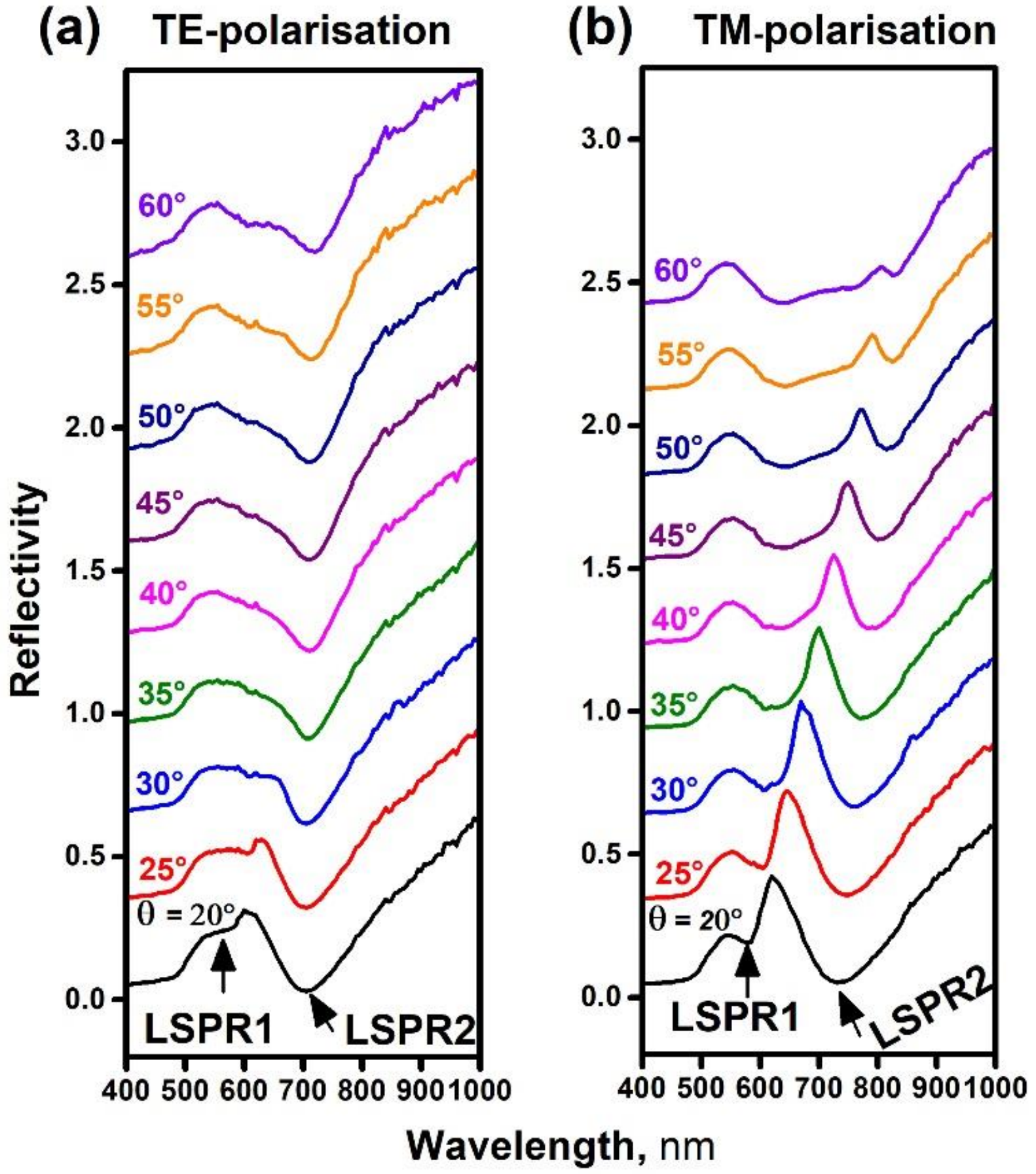

**Fig. 6.** Experimental results of the plasmonic response in a 500 nm diameter AuFON sample, analysed under different angles of incidence (θ) with a constant azimuthal angle (ϕ). Panels (a) and (b) show the reflectivity corresponding to the TE and TM polarisations, respectively. The curves have been shifted vertically for clarity in the presentation of the results.

We now turn to the nanosphere size dependence of the plasmonic response. Figure 7a presents the simulation of the plasmonic response as a function of wavelength for nanostructure diameters between 500 and 900 nm, with 50 nm increments. The simulations were performed with TM polarisation, θ=25°, and ϕ=0°. The result shows a strong dispersion of the plasmonic modes (LSPR1 and LSPR2) towards longer wavelengths (lower energies) as the nanostructure diameter increases, indicating that, as expected, the resonance energy of these modes is sensitive to the nanostructure dimensions. This behaviour is consistent with previous studies in one-dimensional chains of ordered spheres, with variations in the diameter and spacing between spheres [65,66]. The resonance of the LSPR2 mode is more stable against increases in diameter compared to the LSPR1 mode, which weakens for diameters greater than 800 nm. Figure 7b shows the resonance energy as a function of the nanostructure diameter for TM and TE polarisations, θ=25°, and ϕ=0°. It is observed that the resonance energy of the LSPR1 and LSPR2 modes significantly depends on the incident light polarisation, with differences of 21 nm for LSPR1 and 30 nm for LSPR2. These results align relatively well with the experimental results shown in Figure 3, where the resonance differences were 15 and 33 nm for LSPR1 and LSPR2, respectively. This behaviour is attributed to the orientation of the electric field with respect to the metallic surface and its ability to couple to the localised surface plasmon resonances. Although LSPR modes can be excited under both TE and TM polarisations, the excitation efficiency is higher for TM polarisation, particularly under oblique incidence. This difference manifests as slight spectral shifts between the positions of the LSPR1 and LSPR2 modes when comparing both polarisation types. Finally, in Figure 7c, the spatial distribution of the electric field for the LSPR1 and LSPR2 modes in a 500 nm diameter AuFON sample, θ=25°, ϕ=0°, and TM polarisation is presented. The results showed that the electric field was mainly concentrated at the junctions between the nanospheres and their upper surfaces, forming high-intensity regions; this behaviour is consistent across all analysed diameters. These regions exhibit notable spatial confinement of the electric field and correspond to the sites where localised surface plasmon resonance is most effectively excited.

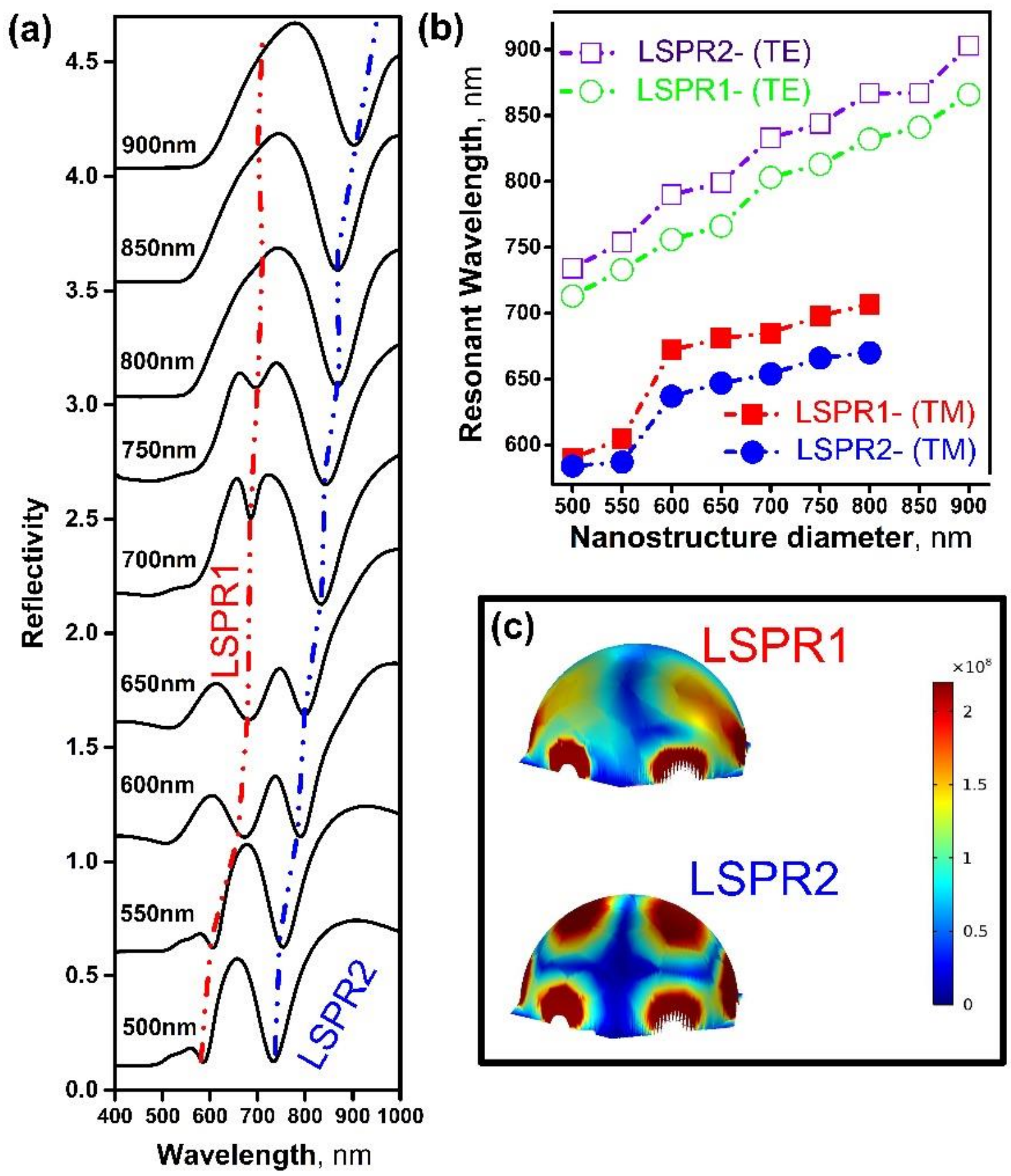

**Fig. 7.** (a) Simulated spectral response for nanostructure diameters ranging from 500 to 900 nm of the AuFON system, with TM polarisation, θ=25°, and ϕ=0°. The curves are vertically shifted by 0.5 for clarity. (b) Resonance wavelength vs. nanostructure diameter for the LSPR1 and LSPR2 modes, with TM and TE polarisations, θ=25°, and ϕ=0°. (c) Spatial distribution of the electric field for the resonant modes LSPR1 and LSPR2, for a 500 nm diameter AuFON sample, θ=25°, ϕ=0° and TM polarisation. In the case of TE polarisation, the spatial distribution was similar to that of TM polarisation.

## 4. Conclusions

In this work, we have studied and optimised the identification of plasmonic resonance peaks in AuFON nanostructured systems. Through detailed characterisation and numerical analysis, we were able to precisely determine the resonance energy of localised plasmonic modes as a function of various parameters, such as the diameter of the nanospheres and the conditions of the incident radiation. In particular, it has been demonstrated that this type of system presents two types of plasmonic resonances, whose spectral positions can be precisely adjusted by varying the light polarisation and the size of the nanostructures. The optimised methodology for identifying these resonance peaks ensures greater efficiency in the amplification of optical signals, which is crucial for technological applications, especially in molecular detection.

Additionally, it has been verified that the resonance of the plasmonic modes does not significantly depend on the azimuthal angle $\phi$, which simplifies optimisation for practical applications. It was found that the LSPR1 and LSPR2 modes exhibit different capacities for electric field intensification, with the LSPR2 mode being the most efficient for amplification applications in molecular detection, due to its stability and higher concentration of the electric field on the nanostructured surface.

The results provide clear criteria for accurately identifying and tuning the spectral position of plasmonic modes in AuFON systems, significantly enhancing their ability to tailor substrates for specific applications, such as molecular detection via SERS. This optimisation facilitates the design of more efficient and reproducible substrates, contributing to the advancement of technologies in molecular diagnostics and environmental monitoring, among other applications.

**Acknowledgments**

The authors declare that there are no acknowledgements to report for this work.

**Funding**

This research received no external funding.

**Conflicts of interest**

The authors declare no conflict of interest.

**Data availability statement**

The experimental and simulation data supporting this study's findings are available from the corresponding author upon reasonable request.

**Author contribution statement**

Luis Alfonso Guerra Hernández: Conceptualization, Methodology, Investigation, Supervision, Writing–Original Draft Preparation, Writing–Review & Editing. Osmar Gil Salas: Methodology, Formal Analysis, Data Curation, Software, Investigation. Jorge Enrique Rueda Parada: Writing–Review & Editing, Conceptualization. Alejandro Fainstein: Conceptualization, Writing–Review & Editing, Investigation, Funding Acquisition. Andrés Alejandro Reynoso: Software, Methodology, Writing–Review & Editing.